\documentclass{elsart}
\usepackage{graphics}
\usepackage{psfrag}
\usepackage{epsfig}
\usepackage{epsf}
\usepackage{float}
\usepackage{amssymb,stmaryrd,latexsym}

\newcommand{\be}{\begin{eqnarray}}
\newcommand{\ee}{\end{eqnarray}}
\usepackage{amsmath}
\usepackage{amssymb}
\usepackage{epsfig}

\begin{document}
\hfill{\small FZJ--IKP(TH)--2005--25}

\begin{frontmatter}
\title{Can one extract the $\pi$-neutron scattering length from $\pi$-deuteron scattering?}

\author{A. Nogga and  C.~Hanhart}

{\small Institut f\"{u}r Kernphysik, Forschungszentrum J\"{u}lich GmbH,}\\ 
{\small D--52425 J\"{u}lich, Germany} \\

\begin{abstract}
\noindent 
We give a prove of evidence that the original power counting by Weinberg can
  be applied to estimate the contributions of the operators contributing to
  the $\pi$-deuteron scattering length. As a consequence, $\pi$-deuteron observables can be
  used to extract neutron amplitudes---in case of $\pi$-deuteron scattering this
  means that the $\pi$-neutron scattering length can be extracted with
  high accuracy.  This result is at variance with recent claims. We discuss
  the origin of this difference.
\end{abstract}

\end{frontmatter}

{\bf 1.} 
In absence of neutron targets, it became common practice to use few--body
nuclei as effective neutron targets. To extract $\pi$-neutron ($\pi$-n)
amplitudes, $\pi$-deuteron ($\pi$-d) scattering has been studied in the past.
This program can be successful only when both the proton observables and the
few--body corrections are known to high accuracy.  As the former can be
measured directly, they do not cause any problem. For the latter the
development of chiral perturbation theory for few--nucleon systems promised a
controlled, model independent, high precision evaluation of the corresponding
amplitudes. This program was put forward in a series of publications, e.g.,
for $\pi$-d scattering (see \cite{BBEMP} and references therein).

\begin{figure}[t!]
\begin{center}
\epsfig{file=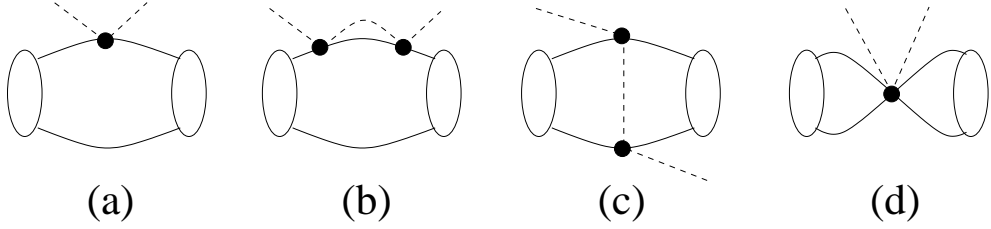, height=3cm, angle=0}
\caption{Typical contributions to $\pi d$ scattering. Diagram $(a)$ and $(b)$
show the tree level and the one--loop contribution to the one--body term, the
pion rescattering contribution is depicted by $(c)$ and diagram $(d)$ shows a
two--nucleon contact term. In this figure solid (dashed) lines denote nucleons
(pions) and ellipses the deuteron wave function.  }
\label{dia}
\end{center}
\end{figure}

All those analyses are based on the conjecture of Ref. \cite{wein} that the
transition operators for reactions on nuclei with external sources can be
constructed perturbatively within chiral perturbation theory. The resulting
operators are then to be convoluted with the appropriate nuclear
wave functions. For this to work it needs to be assumed that the
contribution of few--nucleon counter terms to the transition operators can be
estimated on the basis of na\"{i}ve dimensional analysis. 
If we apply this recipe to $\pi d$ scattering, the leading counter term
(Fig.  \ref{dia} (d))
appears at 5th order---two orders down compared to the leading few body
correction
(Fig.  \ref{dia} (c)).
This was recently confirmed by an explicit calculation of the counter term
contribution assuming natural strength for the transition operator~\cite{rus}.

In contrast to this it was found recently that a logarithmic scale dependence
shows up in the leading few--body correction to $\pi$-d scattering (Fig.
\ref{dia} (c)) that calls for a counter term already at this very order
\cite{gries,rus} (see also \cite{silasmartin}), which would preclude any high
accuracy extraction of $\pi$-n scattering parameters from $\pi$-d data.  This
finding is based on a perturbative treatment of one--pion exchange.

In contrast, in this letter, we demonstrate by an explicit numerical
calculation that the
logarithmic divergence disappears, if we treat the one--pion exchange
non--perturbatively to obtain the wave function.  This explains, why previous
studies basically lead to identical numbers for the leading few--body
correction although very different wave functions were employed (see
discussion in Ref. \cite{BBEMP})\footnote{Please note that in
Ref. \cite{mehensteward}, it was shown that in the deuteron channel pions
should not be treated perturbatively.}. Stated differently, we will show that the
contact term that necessarily arises at next-to-leading order (NLO), 
when pions are treated
perturbatively in the wave function, can be calculated once the pion exchange
is included non--perturbatively in the wave function. This was already
conjectured in Ref. \cite{silasmartin}, but not shown explicitly.

Thus the main goal of our study is to investigate the regulator dependence of
the leading few--body correction. Since we are going to employ wave functions
that contain non--perturbative pion contributions, this study can only be
performed numerically. We will use deuteron wave functions that were
constructed for cut--offs that vary over a wide range
($\Lambda=$ 2--20~fm$^{-1}=$ 400--4000~MeV). 
The procedure of their construction is described
in Ref.~\cite{TNK} and will be briefly reviewed below.  Already in
Ref. \cite{BBEMP} a mild cut--off dependence was reported for calculations
using wave functions with non--perturbative pions, when the regulator was
changed from 500 to 600 MeV. This might either be because of the absence of
the logarithmic divergence due to the wave functions used or simply because
the coefficient in front of the logarithm is accidentally small. Due to the
large range of variation of cut--off values used here we are in the position
to answer this question: we will show that there is no sign of a logarithmic
regulator dependence of the results as soon as the complete wave functions are
used. The consequences of this observation will be discussed in the final
section.

In Ref.~\cite{recoils,lensky}, it was stressed that care has to be taken when
calculating pion reactions on nuclei.
There it was shown that a subtle cancellation pattern exits
 between contributions from loops in
one--body and few--body operators. This has  the effect that the static pion
exchange is an excellent approximation to the exact result for the leading
few--body corrections to $\pi$-d scattering. Therefore here we will focus on
the static exchange only.

\begin{figure}[t!]
\begin{center}
\epsfig{file=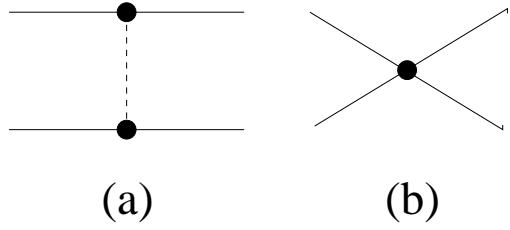, height=3cm, angle=0}
\caption{Contributions to the $NN$ potential at leading order:
the one pion exchange (a) and a momentum independent contact term (b).}
\label{potpic}
\end{center}
\end{figure}

{\bf 2.} In our investigation we use the wave functions constructed as
outlined in Ref. \cite{TNK}. They emerge as a solution of the Schr\"odinger
equation
\begin{equation}
\Psi_\Lambda^\pi(p) = G(p)\int d^3p'V(p,p')f_\Lambda (p,p')
\Psi_\Lambda^\pi (p') \ ,
\label{schr}
\end{equation}
where $G(p)=(-\epsilon-p^2/M)^{-1}$ denotes the two--nucleon propagator with
$\epsilon$ and $M$ for the deuteron binding energy and the nucleon mass,
respectively.
The leading order potential $V(p,p')$ comprises contributions from both the one--pion
exchange as well as a contact term as depicted in Fig. \ref{potpic} (see \cite{weinbergnn}).
As  regulator function we use
\begin{equation}
\label{eq:reg}
f_\Lambda(p,p') = \exp \left( p^4+p'\, ^4 \over \Lambda^4 \right) \ .
\end{equation}
For a given value of the regulator $\Lambda$ the only free parameter is $C$---the
strength of the contact term as depicted in Fig.~\ref{potpic}(b). 
For this study, this parameter was adjusted such that the deuteron binding
energy was reproduced, to exclude 
any dependence of the results on an incorrect asymptotic behavior of the 
deuteron wave function. We checked that the description of the phase shifts in 
the $^3$S$_1$-$^3$D$_1$ channel is comparable to the one obtained 
in \cite{TNK}. This numerical study can only be conclusive, when we cover 
a wide range of cut--offs. We decided to use values of
$\Lambda$ between 2 and 20 fm$^{-1}$ (400--4000~MeV). 
This range starts below the 
chiral symmetry breaking scale of $\Lambda_\chi \approx 1000-1200$~MeV  and 
extends to values larger by a factor of 4. In this range, we also observe 
the appearance of spurious bound states in the $^3$S$_1$-$^3$D$_1$
channel. However, their energies are large and, therefore, these bound states 
should not affect any low energy physics. 
 
For comparison we also prepared a series of wave functions from only contact
$NN$ interactions (thus omitting diagram \ref{potpic}(a) in the
potential). These wave functions are denoted by $\Psi_\Lambda^{no \ \pi}(p)$ in
what follows. Again we impose a regulator as given in Eq.~(\ref{eq:reg}) 
in the Schr\"odinger equation. 

For completeness, we summarize the binding energy results and some 
wave function properties for both series of wave functions in 
Table~\ref{tab:piwf} and \ref{tab:contwf}. 

\begin{table}[t]
\begin{center}\begin{tabular}{r|rrrrrrr}
$\Lambda$  & $E_0$  &  $T$    & $P_D$ &  $A_S$ 
 & $\eta$   &  $r_d$  & $Q_d$ \cr
\hline
2  & 2.225 &  28.91 & 5.24  & 0.839 & 0.030 & 1.889 & 0.3005  \cr
4  & 2.225 &  45.48 & 8.23  & 0.866 & 0.027 & 1.933 & 0.2827  \cr
6  & 2.224 &  62.33 & 6.94  & 0.866 & 0.025 & 1.932  & 0.2704 \cr
8  & 2.225 &  75.95 & 6.76  & 0.864 & 0.026 & 1.926  & 0.2676 \cr
12  & 2.227 & 85.80 & 7.14 & 0.864 & 0.026 &  1.925 & 0.2675 \cr
 16.5  & 2.214 & 102.50 & 7.08 & 0.862 & 0.026 & 1.929 & 0.2676 \cr
 20    & 2.210  & 115.07  & 7.07 & 0.861 & 0.026 & 1.929 & 0.2675 \cr
\hline
Expt. & 2.225 & ---  & ---  & 0.8846 & 0.0256  & 1.9671  & 0.2859  \end{tabular}
\end{center}\caption{\label{tab:piwf} Summary of some deuteron properties
  obtained from $\Psi_\Lambda^\pi$---the wave functions with non--perturbative 
one--pion exchange for various cutoffs. Here, the cut-off $\Lambda$ is given in 
fm$^{-1}$, the binding energy and kinetic energy $E_0$ and $T$ in MeV, 
the asymptotic $S$-state normalization $A_S$ is in fm$^{-1/2}$, 
the point nucleon radius in fm, and the quadrupol moment in fm$^2$. 
$\eta$ is the ratio of 
the asymptotic $S$- and $D$-state normalization.
}\end{table}

\begin{table}[t]
\begin{center}\begin{tabular}{r|rrrr}
$\Lambda$ & $E_0$  &  $T$  &  $A_S$  &
    $r_d$  \cr
\hline
2  & 2.225 &  32.60 & 0.76 & 1.728   \cr
4  & 2.225 &  69.59 & 0.72 & 1.617 \cr
6  & 2.225 &  106.76 & 0.71 & 1.585   \cr
8  & 2.225 &  143.99& 0.70& 1.569  \cr 
12  & 2.225 & 218.51 & 0.69 &  1.555 \cr
 16   & 2.225 & 293.04 & 0.69 & 1.547  \cr
 20    & 2.225  & 367.59  & 0.69 & 1.543 \cr
\hline
Expt. & 2.225 & ---  & 0.8846 & 1.9671  \end{tabular}
\end{center}\caption{\label{tab:contwf} Summary of deuteron properties
  obtained from the wave functions $\Psi_\Lambda^{no \ \pi}$---where only a contact 
interaction was used in the potential---for various cutoffs. The notation is the same 
as in Table~\ref{tab:piwf}. }\end{table}

{\bf 3.} Let us now turn to the calculation of the leading few--body correction
to the $\pi d$ scattering length. As stated in the introduction we will
exclusively focus on the static contribution. The corresponding expression
reads
\begin{equation}
a^{(static)}=-\xi \int
d^3pd^3q\Psi_\Lambda^\kappa(\vec p - \vec q)^\dagger \frac{1}{\vec q\, ^2}\Psi_\Lambda^\kappa(\vec p) \ ,
\label{stat}
\end{equation}
where $\xi=m_\pi^2/(32\pi^4 f_\pi^4\left (1+ m_{\pi}/(2M_N) \right
))$. Clearly, no physical quantity can be regulator dependent. Na\"{i}ve dimensional 
analysis does not require a two--body counter term in the same order 
as this first three--body correction. Thus, studying
the $\Lambda$ dependence of the given integral tells, whether a counter 
term of na\"{i}vely higher orders is needed in conjuction with this few--body 
correction. 
We evaluated numerically the integral in
Eq.~(\ref{stat}) using both, the wave functions from the full LO $NN$ potential
($\kappa =\pi$) and those from only the point interaction ($\kappa =no \
\pi$), as described in the previous section. 

\begin{figure}[t!]
\begin{center}
\epsfig{file=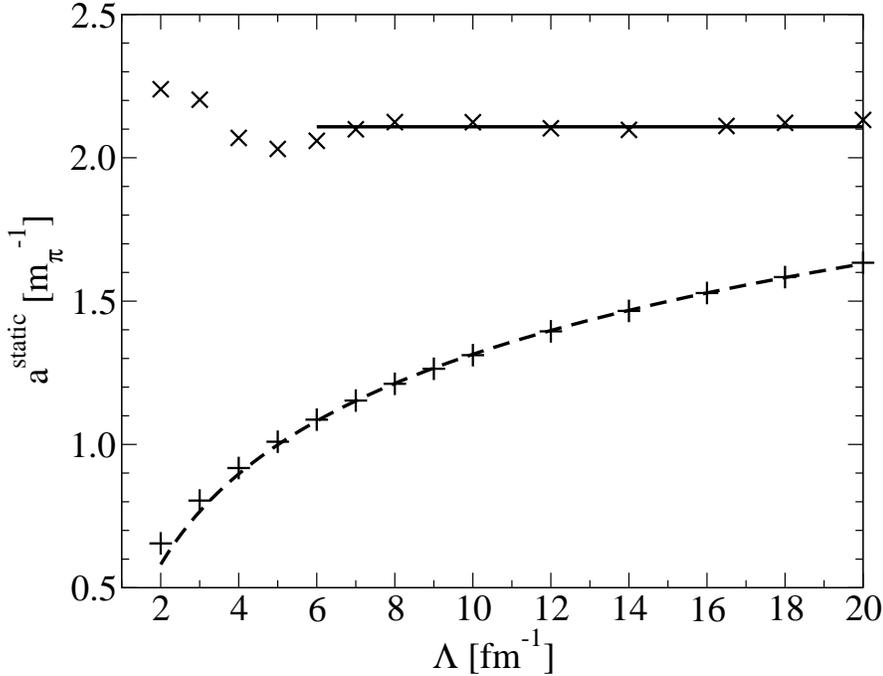, height=10cm, angle=0}
\caption{Results for the $\Lambda$ dependence of the leading few--body
  correction from Eq. (\protect\ref{stat}).  The $'x'$ symbols show the results of the
  numerical evaluation of the integral using the full wave functions
  $\Psi_\Lambda^{\pi}$, whereas the $'+'$ symbols shows the result for wave
  functions with only point interactions $\Psi_\Lambda^{no \ \pi}$, downscaled
  by a factor of 4. The dashed line is a fit to the latter of the form $A+B\ln
  (\Lambda)$, as described in the text. The solid line shows the fit of a
  constant to the former for values of the cut--off larger than 6 fm$^{-1}$.}
\label{results}
\end{center}
\end{figure}

The results are shown in Fig. \ref{results}. Here the x symbols emerged from
the calculation with the wave functions $\Psi_\Lambda^\pi$---where the
non--perturbative pion exchange was included---whereas the plus symbols stem
from the calculation using $\Psi_\Lambda^{no \ \pi}$---that does not have any
pion exchange in the wave function. For the latter, our results clearly show
the $\ln(\Lambda)$ behavior in accordance with the findings of
Refs. \cite{gries,rus,silasmartin}. To show that explicitly, we fitted a
logarithm to our results, which is also shown in the figure. The perfect
agreement shows that we can recover the previous results numerically in our
cut--off range. However, the calculation using the full wave functions shows
almost no regulator dependence at all for cut-offs above
$\Lambda_\chi$\footnote{Note that we observe a mild regulator dependence of
  the integral for regulators below $\Lambda_\chi$, in line with the findings
  reported in Ref. \cite{BBEMP}.}. 
Thus,
as soon as the pion exchange is included non--perturbatively in the wave
functions, no counter term is needed at the order of the leading few--body
corrections to absorb the regulator dependence of the pion exchange
contribution.  For a numerical comparison with previous work, we fitted a
constant to our results for $\Lambda \ge 6$~fm$^{-1}$. In this way, we obtain
for the static three-body contribution $a^{static} = 2.12 $~m$_\pi^{-1}$,
which is in good agreement with the previous calculations \cite{BBEMP}.

We also checked that there is no unnatural enhancement of the leading $\pi
NN\to\pi NN$ counter term due to the wave function at small distances: the
contribution of this term to the scattering length---from an explicit
evaluation of diagram (d) in Fig. \ref{dia}---was in line with the counting,
when the transition operator was assumed to be of natural strength
\cite{friar}. The same observation was also made in Ref. \cite{rus}.
This shows that as soon as non--perturbative pions are included
in the construction of the wave function, the wave function at the origin
assumes natural values. Combining this finding with the observation made above
that there is no regulator dependence of the leading few--body correction,
there is no reason to change the original Weinberg counting.

\begin{figure}[t!]
\begin{center}
\epsfig{file={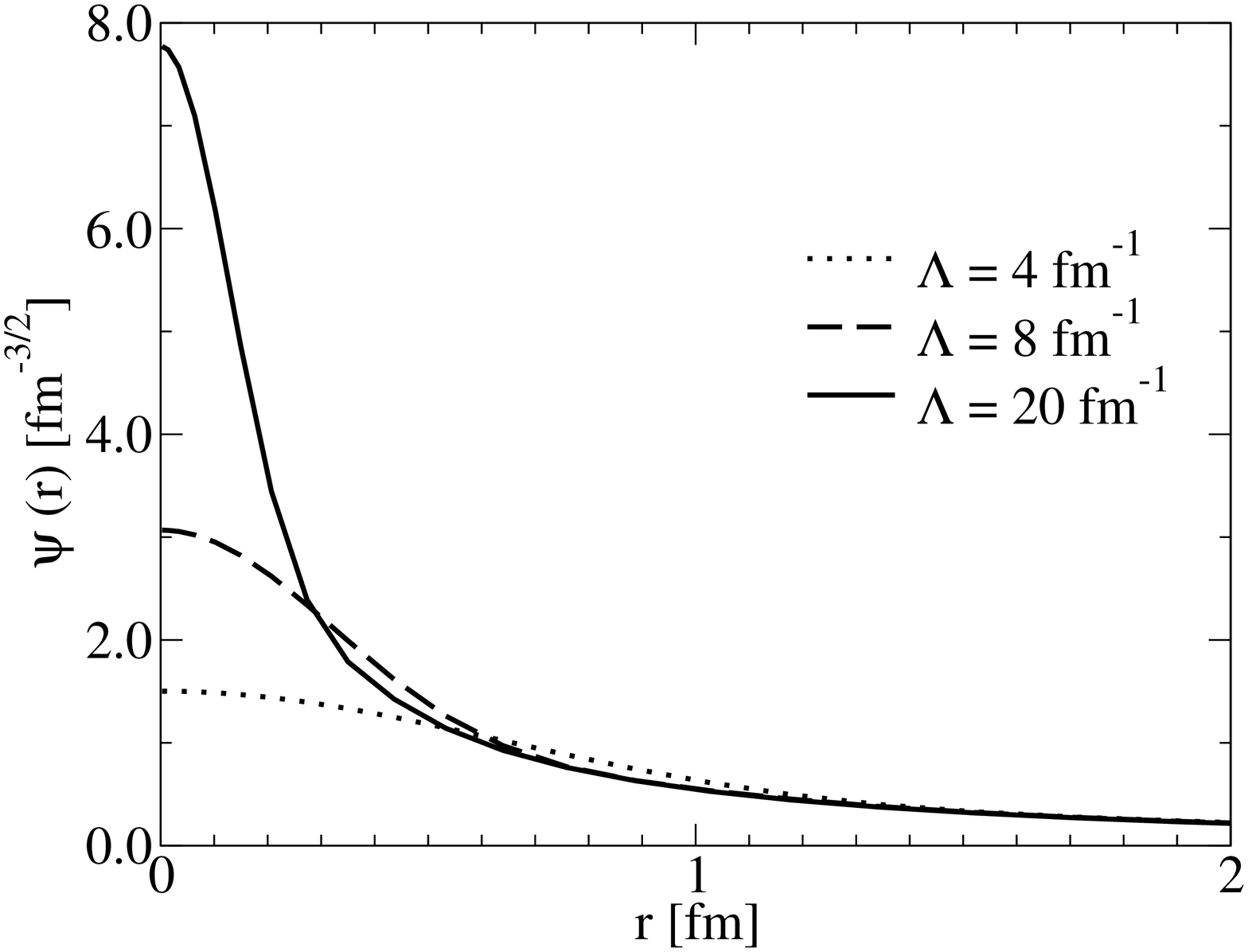}, height=9cm, angle=0}
\epsfig{file={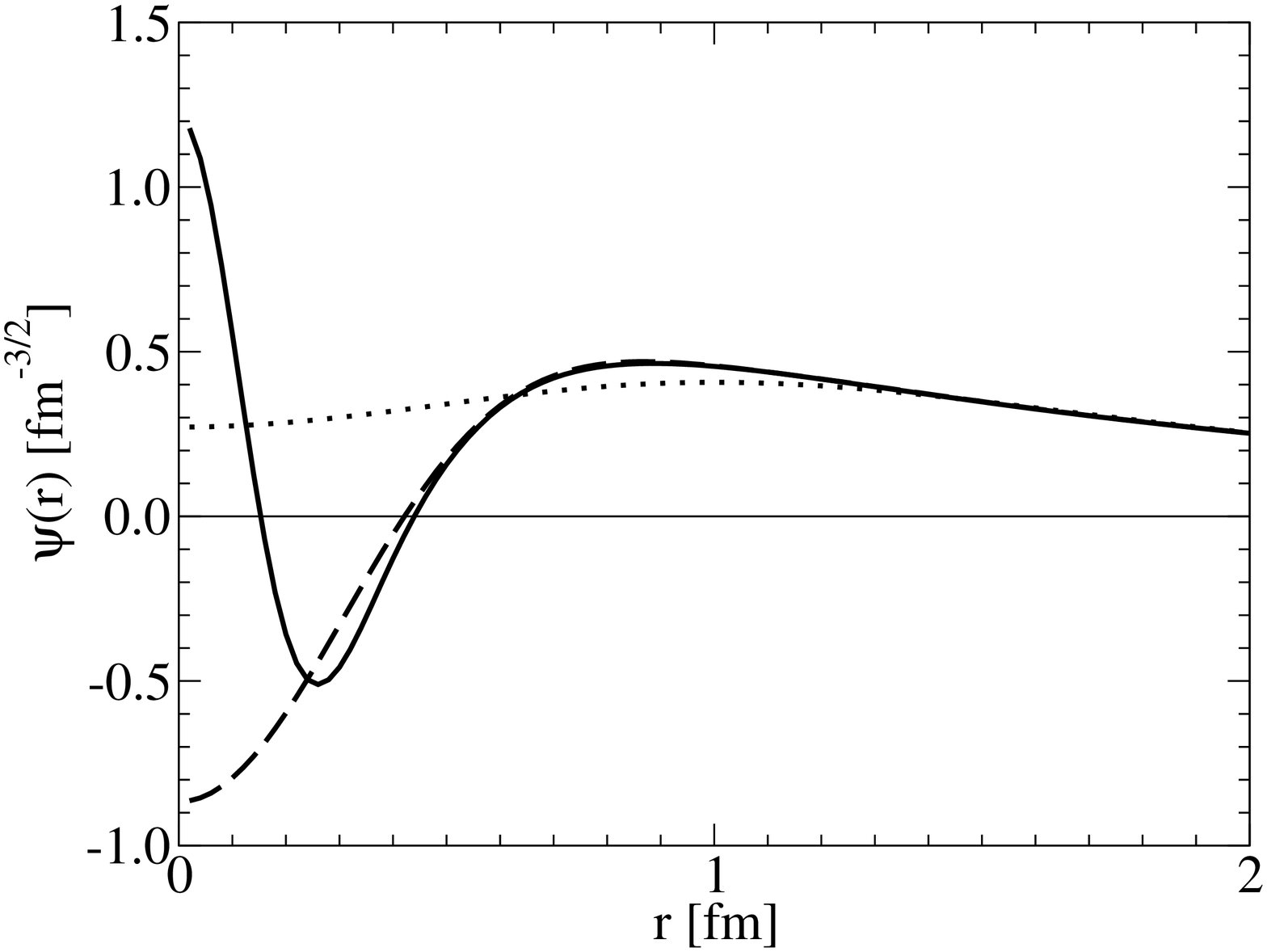}, height=9cm, angle=0}
\caption{ The upper panel shows $\Psi_\Lambda^{no\  \pi}$ and the lower one 
 the s-wave of $\Psi_\Lambda^{\pi}$
 for various values of $\Lambda$. Note the different scales of the figures.}
\label{fig:wf}
\end{center}
\end{figure}

We conjecture that the full wave function is driven not only by the the
binding momentum of the deuteron, as $\Psi_\Lambda^{no \ \pi}$, but also by a
second scale, probably $m_\pi$ or $f_\pi$.  The numerical calculation
does not easily allow to identify such a scale in the wave functions. For
illustration, we compare the two types of wave functions in $r$-space in
Fig.~\ref{fig:wf}.  For $\Psi_\Lambda^\pi$, we see that the wave functions
converge for radii larger than $R=0.7$~fm. 
For smaller $r$, the wave functions are not unique in our range 
of cutoffs and show
different numbers of nodes coming from the spurious bound states.  They
remain, however, small for these distances. This is different for the
point--like wave functions, which, for $r=0$ increase linearly with the 
cut-off.  We conclude that the
finite range of the pion--exchange regularizes the wave functions in the sense
that they remain small for small distances independent of the cut--off and oscillate. 
Both effects reduce the contribution of the small distance behavior to the scattering 
length. A more systematic insight would obviously be highly welcome.

{\bf 4.} To summarize, as soon as a finite range potential is used in the
construction of the deuteron wave functions, the matrix element for the
leading few--body correction to the $\pi$-d scattering length gets independent
of the regulator. In particular, the logarithmic divergence that emerges in
case of wave functions from a point--like $NN$ potential disappears. We have
demonstrated this in a numerical study using wave functions constructed for a
large variety of cut--offs. Our numerical 
results extend to cut-offs, which are much larger than the chiral symmetry 
breaking scale. For the same range of cut-offs, we numerically 
observe the expected logarithmic divergence for the case 
of $\psi^{no \ \pi}$. Therefore, we are confident that the range of 
cut-offs is large enough to take our result as a prove of evidence that
no counter term is needed at the same order where the leading few--body
correction appears. This is the precondition to allow for a high accuracy
extraction of the $\pi$-n scattering length from $\pi$-d data. 

Please note that the present discussion has large implications also for the
investigations of other reactions. Based on our findings the calculations for
$\gamma d\to \pi^0 d$ \cite{kbl,krebs}, $\pi {^3}$He$\to \pi {^3}$He
\cite{baru}, $\pi^- d\to \gamma nn$ \cite{garde}, and $\gamma d\to \pi^+ nn$
\cite{lensky} indeed were performed with the accuracy as given in the
publications. 

Finally, we note that the observation, that the one--pion exchange, which
leads to a finite range interaction, changes the divergence structure of the
theory, is not unique to $\pi$-d scattering. The probably most famous other
example in this context is the three--nucleon system in the spin $1/2$
channel: as long as only point--like two body interactions are included, a
three body counter term needs to be promoted to leading order
\cite{HBvK}. However, as soon as there is a non--perturbative pion--exchange
included, this is no longer necessary \cite{TNK}.

{\bf Acknowledgment}

We thank Akaki Rusetsky and Ulf--G. Mei\ss ner for stimulating discussions during the 
preparation of this letter. This work was supported by
DFG grant no. 436 RUS 113/820/0-1(R). The numerical calculations were 
performed on the JUMP cluster of the NIC in J\"ulich, Germany.

\end{document}